\documentclass[twocolumn,preprintnumbers,amsmath,amssymb]{revtex4}

\usepackage{graphicx}  
\begin{document}

\title
{Device model for pixelless infrared image up-converters 
based on polycrystalline graphene heterostructures}

\author{V.~Ryzhii$^{1,2,3}$,  M.~S.~Shur$^4$,  M.~Ryzhii$^5$,  V.~E.~Karasik$^3$, 
and  T.~Otsuji$^1$}
\address{
$^1$ Research Institute of Electrical Communication, Tohoku University,
 Sendai 980-8577, Japan\\
$^2$ Institute of Ultra High Frequency Semiconductor Electronics of RAS,\\
 Moscow 117105, Russia\\
$^3$ Center for Photonics and Infrared Engineering, Bauman Moscow State Technical University, Moscow 111005, Russia\\
$^4$ Department of Electrical, Computer, and Systems Engineering, Rensselaer Polytechnic Institute, Troy, New York 12180, USA\\
$^5$ Department of Computer Science and Engineering, University of Aizu, 
Aizu-Wakamatsu 965-8580, Japan}
%


\begin{abstract}
We develop a device model for pixelless converters of far/mid-infrared radiation (FIR/MIR) images  into 
near-infrared/visible (NIR/VIR) images. These converters use polycrystalline graphene layers (PGLs) immersed in 
the van der Waals (vdW) materials integrated with light emitting diode (LED). The PGL serves as an element of the PGL infrared photodetector (PGLIP) sensitive to the incoming FIR/MIR due to the interband absorption.
The spatially non-uniform photocurrent generated in the PGLIP repeats (mimics) the non-uniform distribution (image) created by the incident FIR/MIR. The injection of the nonuniform photocurrent into the LED active layer results in the nonuniform NIR/VIR image reproducing the FIR/MIR image. The PGL and the entire layer structure are not deliberately partitioned into pixels. We analyze the characteristics of such pixelless PGLIP-LED up-converters
and show that their image contrast transfer function and the up-conversion efficiency depend on the PGL lateral resistivity. The up-converter exhibits  high photoconductive gain and conversion efficiency when the lateral resistivity is sufficiently high. Several teams have successfully demonstrated the large area PGLs with the resistivities varying in a wide range. Such layers can be used in the pixelless PGLIP-LED image up-converters.    
The PGLIP-LED image up-converters can substantially surpass the image up-converters based on the quantum-well infrared photodetector (QWIP) integrated with the LED. These advantages are due to the use of the interband FIR/NIR absorption and a high photoconductive gain in the GLIPs.
\end{abstract}

\maketitle

\begin{figure*}[t]
\centering
\includegraphics[width=10.0cm]{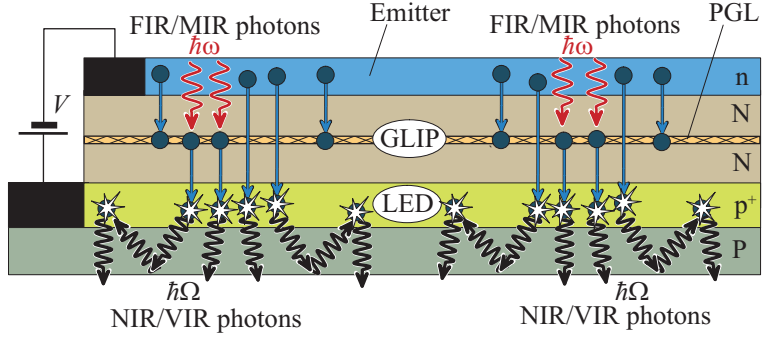}
\caption{ Schematic view of (a) the PGLIP-LED up-converter structure. Wavy arrows correspond to the
incident photons (with the energy $\hbar\omega$) and the photons generated in the LED part (with the energy $\hbar\Omega$). Arrows indicate passes of the electrons
injected from the emitter n-emitter layer and those  excited from the GL by FIR/MIR  photons.}
\end{figure*}

\begin{figure*}[t]
\centering
\includegraphics[width=15.0cm]{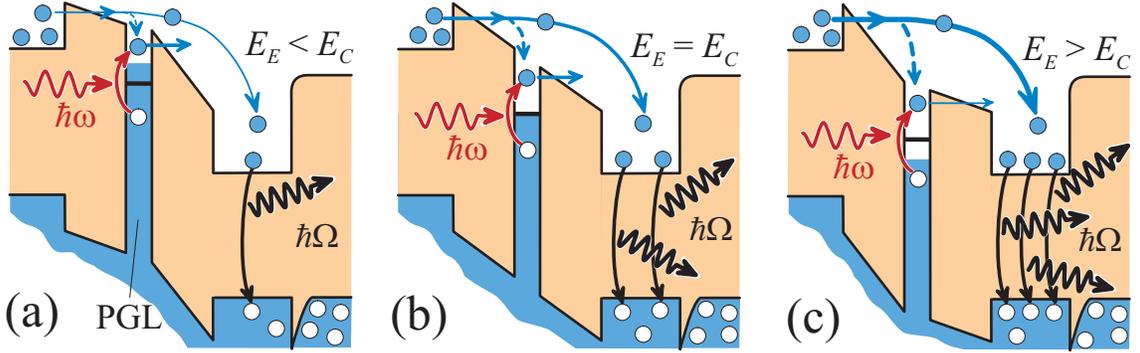}
\caption{ Schematic view of the PGLIP-LED up-converter band diagrams with  the emitter electric field $E_E$ (a) smaller than the collector field $E_C$,  (b)  $E_E = E_C$, and  (c)  $E_E > E_C $, respectively. Dashed arrows correspond to the electron capture into the GLs.}
\end{figure*}

\section{Introduction}  

The main problem in the transformation of far-infrared radiation (FIR), mid-infrared radiation (MIR), or near- infrared radiation (NIR) images, into visible (VIR) or even ultraviolet images is the availability of the pertinent detector technology. 
Despite tremendous success associated with CCD and CMOS digital technology, imaging at relatively long wavelengths where silicon is "blind", is very complicated and expensive~\cite{1,2}.
Therefore,  the demand for practical devices effectively converting
FIR, MIR, and NIR images to VIR images is very strong. Different approaches have been  explored, including  thermal imaging, nonlinear up-conversion  and photochemical up-conversion based on sensitized  triplet-triplet annihilation, and others~\cite{2,3,4}. In particular, 
 the integration of the  quantum-well infrared photodetectors (QWIPs) with the light-emitting diodes (LEDs) for the image up-conversion was proposed and implemented almost two decades ago~\cite{5,6,7,8,9,10,11}. However, despite reasonable characteristics of the QWIP-LED image up-converters, they have not found wide applications
because of the  incline incidence requirement (or the necessity to use special radiation couplers), a relatively low conversion efficiency (due to a relatively low {\it intersubband} radiation absorption and the absence
of the photoelectric gain resulting in a modest contrast transfer). Technologically, the realization of effective QWIP-LED up-converter requires the formation of large area multiple-QW heterostructures.
Some drawbacks of the QWIP-LED image up-converters might be eliminated in the image up-converters based on 
 the  integration of quantum-dot infrared photodetectors (QDIPs)~\cite{16} and   QD- or QW-LEDs as was proposed in Ref.~\cite{17}. But this idea was not realized yet, although the lamp (pixell) QD-based up-converters were recently reported~\cite{18,19,20}.

Recently, we proposed to use the graphene-layer infrared photodetectors (GLIPs) integrated with the light-emitting diodes (LEDs) for the photon energy up-conversion leading to the transformation of far/mid-infrared
(FIR/NIR) signals into near-infrared/visible (NIR/VIR) signals~\cite{12}. In such GLIP-LED up-converters, the photocurrent produced in the GLIP part of the device due to the FIR/MIR {\it interband} absorption~\cite{13,14,15}  is injected into a LED resulting  
in the emission of NIR/VIR. The GLIP-LED elements can form  pixels of the system (which consists of an array of such pixels) up-converting
the FIR/MIR images. In this paper, we show that the GLIPs with the large area macroscopically uniform and sufficiently resistive polycrystalline GLs  (PGLs) (not intentially partitioned into  pixels)  integrated with the large area LEDs can up-convert the FIR/MIR images.
We develop the device model for the pixelless PGLIP-LED image up-converters and evaluate
 their characteristics.  The operation of the   PGLIP-LED up-converters is associated
  with the injection of  the spatially nonuniform photocurrent
produced in the GLIP part of the device by   the spatially nonuniform FIR/MIR (FIR/MIR image)   into its  LED
part resulting 
in the emission of the spatially nonuniform NIR/VIR (i.e.,NIR/VIR image). 
The pixelless PGLIP-LED image up-converters  can be implemented 
  in the heterostructures with the PGL and the barrier layers
made of different materials, in particular, the so-called van der Waals (vdW) materials~\cite{21,22,23,24,25,26,27} (hBN, WS$_2$, WSe$_2$, and many others) and using these materials for 
the LED part of such devices~\cite{28,29,30,31}.  A weak inter-layer bonding enables 
effective  stacking of these layers with different  lattice constants.
The pixelless PGLIP-LED image up-converters can surpass  the pixeless QWIP-LED image up-converters due to:\\
 (i) the GL (and PGL)  sensitivity  to the normally incident input FIR/MIR~\cite{32} because of the use of the interband transitions (avoiding the need  for FIR/MIR  coupling structures);\\
  (ii) a higher probability of the direct or followed by tunneling electron photoexcitation from the GLs (than that from QWs) into the continuum states above the inter-GL barriers~\cite{32,33,34,35};\\
   (iii) the photoconductive gain due to
the  possibility of  nonuniform lateral potential distribution formation in the PGLs with relatively high lateral resistivity  under the nonuniform incident radiation (such a gain occurs due to a low probability of the capture of the electrons into the GL~\cite{36} and can provide substantially higher contrast of the output images and elevated up-conversion efficiency);\\
(iv) easy fabrication due to the robust technology of large size formation of PGLs~\cite{37,38,39,40,41}
with relatively low conductivity due to their polycrystalline nature (with the scattering of charge carriers at grain boundaries degrading their performance relative to exfoliated, single-crystal graphene) and as well as due to other types of disorder~\cite{41,42,43,44,45,46,47}. 

In contrast to the QWIP-LED image up-converters in which multiple-QW structures are indispencible~\cite{5,8}, the PGLIP-LED devices can comprise a single PGL.

These advantages of the pixelless PGLIP-LED image up-converters should  stimulate their  implementation and use in different applications.

\section{Device structure and model}

Figure~1 shows the PGLIP-LED device structure with a single PGL sandwiched by the N-barrier layers and with the top emitter n-layer. The structure comprises also the p$^+$-layer (on the P-type substrate), which serves as the active region of the LED part of the device. Figures 2(a) - 2(c) show the band diagrams corresponding to different 
 electric field in the emitter barrier layer  $E_E$ and in the collector barrier layer $E_C$. The PGLIP part of the device structure under consideration  is
somewhat different from those studied in Refs.~\cite{12,13,14,15}, where
 the emitter n-layer is assumed to be a GL. 

Under the bias voltage applied between the n-emitter  and p$^+$-collector layer (serving as the LED active region), the electron tunneling  through the triangular barrier
provides the electron injection from the emitter to the barrier layer between the  n-emitter region and the PGL. A portion of the injected electrons crosses the PGL and enters to the collector barrier layer and
than to the p$^{+}$-layer. The electron tunneling or thermionic emission from the PGL also contribute to the net current collected by the p$+$-layer. The incident FIR/MIR spatially nonuniform in the device plane
generates the spatially nonuniform electron photocurrent from the PGL. It is associated with the electrons photoexcited in the PGL from its valence band into the conduction band (see Figs.~1 and 2) which go to the barrier layer either directly or after the tunneling through the triangular barrier between the PGL and the collector barrier layer
(depending on the FIR/MIR photon energy $\hbar\omega$ and the height $\Delta_{GL}$ of the barrier between the PGL and the collector barrier layer). The spatial distribution of the photocurrent from the PGL repeats the spatial distribution of the incident FIR/MIR.
The photoexcitation of the PGL leads to the deviation of its potential $\varPhi_{GL}$ from its dark value 
$\varPhi_0^{dark}$. If the PGL lateral resistivity is relatively low (as in sufficiently perfect and/or doped GLs), $\varPhi_{GL}$ is virtually independent of the lateral coordinates, so that the variation of the photocurrent  injected    
from the emitter is uniform as well. Thus, in such a case, the nonuniform irradiation leads to the nonuniform current generated solely from the PGL, whereas the net photocurrent is produced by both the PGL and the emitter. Similar situation occurs in the pixelless QWIP-LED image up-converters due to low QW lateral resistivity. The latter can not normally be made sufficiently high because of the necessity
to maintain relatively high electron concentration (doping) in the near emitter QW to provide a sufficient intersubband absorption and photoemission. 
In contrast, in the PGLIP-LED devices, the lateral resistivity of the PGL  can be so high that the nonuniform distributions of the photogenerated holes (left in the PGL after the escape of the  photoelectons)
do not manage to relax. Hence in this case, the PGL electric potential spatial distribution  becomes 
similar to that of the incident radiation. This results in the nonuniform density of the photocurrent
emitted not only from the PGL but also injected from the n-emitter.
As a result, the spatially nonuniform component of the net current stimulated by the incident FIR/MIR 
can be larger than the component associated with the photoemission from PGL solely. In other words, the effect of photoconductive gain amplifies not only spatially uniform currents (the dark current and the current generated by the averaged component of the incident FIR/MIR intensity $I_{\omega,0} = \langle I_{\omega} \rangle$) but the "image" component as well.

The PGLIP-LED image up-converter model accounts for the main processes responsible for the device operation, namely,
the electron photoemission from the PGL (direct and followed by tunneling), capture of the electrons injected from the emitter into the PGL,
processes of the PGL  lateral conductivity,   injection of the photocurrent to the LED active layer,  and the lateral electron propagation due to the diffusion and the  reabsorption (recycling) of the NIR/VIR photons trapped in this layer.
The main feature of the device under consideration is the use of large area polycrytalline GL as a photosensitive element
with decreased dc conductivity.

In the absence of irradiation, the densities of the electron tunneling current from the emitter and the current of the electrons photoescaped from the  PGL,
$j_E$ and $j_{GL}$, respectively, can be presented as
$j_E = j_E^{max}\exp(- E_{E}^{tunn}/E_E)$ and $ j_{GL} = j_{GL}^{max}\exp(- E_{GL}^{tunn}/E_C)$.
Here $j_E^{max}$ and $j_{GL}^{max}$ are the maximum electron current densities, which can be extracted from the emitter n-layer and the PGL. These quantities are determined by the doping and the electron try-to-escape times. The characteristic tunneling fields for the near-equilibrium electrons in the n-emitter and for the photoexcided electronsin the PGL are equal to $E_{E}^{tunn} = 4\sqrt{2m}\Delta_E^{3/2}/3e\hbar$ and $E_{GL}^{tunn} = 4\sqrt{2m}\Delta_{GL}^{3/2}/3e\hbar$~\cite{48}, respectively, where $\Delta_E$ and $\Delta_{GL}$ are
the electron activation energies in the n-emitter layer and the GL, $m$  is the electron effective mass in the barrier layers, $e$ is the electron charge, and $\hbar$ is the Planck constant. The emitter and collector fields $E_E$ and $E_C$ satisfy the equation $E_EW_E + E_CW_C = V$, were $W_E$ and $W_C$ are the thicknesses
of the barrier layers and $V$ is the bias voltage. In the following, to avoid to cumbersome formulas we, for simplicity, set $W_E = W_C = W$ and 
$j_E^{max} = j_{GL}^{max} = j^{max}$.

Equalizing the capture rate of the injected electrons crossing the PGL  into the latter $j_Ep/e$, where $p < 1$
or $p \ll 1$
is the capture probability (capture parameter) of an electron crossing the PGL into it~\cite{5,7,8,9,11,12,13,14,15}, 
and the rate of the electron tunneling escape from the GL $j_{GL}/e$, one can find in the case of the undoped GL, which will be primarily considered in the following, the condition

\begin{equation}
E_E = E_C = \frac{V}{2W}
\end{equation}\label{eq1}
is achieved at $V = V_0$ with

\begin{equation}\label{eq2}
V_0 = \frac{2W(E_{GL}^{tunn} -  E_{E}^{tunn})}{\ln (j_{GL}^{max}/pj_E^{max})}.
\end{equation}

In the situation under consideration, the surface charge in the PGL $\Sigma = 0$, so that the PGL Fermi level
coincides with the Dirac point, carrier density is minimized, that promotes an elevated GL resistivity.
Such a situation can take place when $E_{GL}^{tunn} > E_E^{tunn}$, i.e., 
when $\Delta_E < \Delta_{GL}$. The latter inequality implies that $\Delta_E = \chi_E - \chi_B - \varepsilon_F <  \Delta_{GL} =\chi_{GL} - \chi_B $, where $\varepsilon_F$ is the electron Fermi in the emitter. Here $\chi_E - \chi_B$ and $\chi_{GL} - \chi_B$ are the differences
between the electron affinities of the emitter material ($\chi_E$) and of the PGL ($\chi_{GL}$)
and that of the barrier material $\chi_B$. Hence, the structure materials and the emitter doping should be chosen in a such a way
that $\chi_E  > \chi_{GL} > \chi_B$ and  $\varepsilon_F > \chi_E  > \chi_{GL}$.

The deviation of $V$ from $V_0$ leads to $E_E > E_C$ or $E_E < E_C$
and to the formation of the excess electron or hole charges in the undoped GL. Usually the latter can result in a marked drop of the GL resistivity. If the PGL is doped, the appropriate choice
of the bias voltage $V = V_0^{doped} \neq V_0$, can decrease the carrier density and, hence, increase the GL resistivity. In this case,
$E_E \neq E_C$, although the PGLIP-LED characteristics can be found analogously.

The consequences of the departure of the GL electron-hole system from the Dirac point  will be discussed below.

To provide an effective injection of the electrons from the GLIP part to the p-layer in the LED part and
NIR/VIR emission from the latter, the following two conditions should be fulfilled: (1) absence of the barrier
at the p-layer and (2)  sufficiently large band gap in the latter layer (to secure emission
of the NIR/VIR photons).
The first  condition requires $\chi_B \leq \chi_{LED} $.

\section{Current output from the GLIP}

The intensity $I_{\omega}^{in}$ of the incident FIR/MIR with the frequency $\omega$ and the variation of the GL potential caused by irradiation $\varPhi_{GL}$
comprise the spatially averaged and spatially nonuniform (in the in-plane $x$-direction) components:

\begin{equation}\label{eq3}
I_{\omega}^{in} = I_{\omega,0}^{in} + I_{\omega, q}^{in}\cos qx, \,\,  \varPhi_{GL} = \varPhi_{GL,0} + \varPhi_{GL,q}\cos qx.
\end{equation}
Here $q$ is the wavenumber characterizing the scale of the image details. The components o of the injected current density induced by the incident FIR/MIR (photocurrent density) are given by

\begin{equation}\label{eq4}
j_{E,0} =   \sigma_E\varphi_0/W,  \qquad  j_{E,q} =   \sigma_E\varphi_q/W,     
\end{equation}
where $\sigma_E = dj_E/dE|_{E = E_E} = j_{E}^{max}\exp(-E_E^{tunn}/E_E) (E_E^{tunn}/E_E^2)$ is the differential conductance of the emitter. 
The spatially uniform components  $\varphi_0$ and $j_{E,0}$ can be found accounting for the balance of the electron captured into and photoescaped from the GL. As a result,

\begin{equation}\label{eq5}
j_{E,0} = \frac{4\pi\,e\alpha\theta_{\omega}}{p(\sqrt{\kappa} +1)^2} I_{\omega,0}^{in}
\end{equation}
with the quantity

\begin{equation}\label{eq6}
\theta_{\omega} =\frac{1}{1 + \displaystyle\frac{\tau_{esc}}{\tau_{relax}}
\exp\biggl(\frac{\eta_{\omega}^{3/2}E_{GL}^{tunn}}{E_C}\biggr)}
\end{equation}
describing the dependence of the electron photoescape on the FIR/IR photon energy $\hbar\omega$~\cite{12,13,14,15}, 
$\eta_{\omega} = (\Delta_{GL} - \hbar\omega/2)/\Delta$,
$\alpha \simeq 1/137$ is the fine structure constant and $\sqrt{\kappa}$
is the barrier material refractive index.

Taking into account the spreading of the holes photogenerated  in the PGL due to the lateral conductivity of the latter, the spatially nonuniform components of the PGL potential $\varPhi_{GL,q}$ can be derived using the following equation (the continuity equation):

\begin{equation}\label{eq7}
\frac{d^2\varPhi_{GL,q}}{d x^2} - Q_{GL}^2\,\,\varPhi_{GL,q} = \frac{4\pi\alpha\theta_{\omega}\rho_{GL}}{(\sqrt{\kappa} +1)^2}
I_{\omega,q}^{in}\cos qx,
\end{equation}
Here $Q_{GL} = \sqrt{p\sigma_E\rho_{GL}/W}$ is the parameter characterizing the lateral spreading of the GL potential  and $\rho_{GL}$ is the GL resistivity.

Equations~(4) and (7) yield

\begin{equation}\label{eq8}
\varPhi_{GL,q} = - \frac{\pi\alpha\theta_{\omega}\rho_{GL}}{(\sqrt{\kappa} +1)^2}\frac{I_{\omega, q}^{in}\cos qx}{(q^2 + Q_{GL}^2)},
\end{equation}
so that the spatially nonuniform component  of the electron current density from the emitter reads

\begin{equation}\label{eq9}
j_{E,q} = \frac{\sigma_E\rho_{GL}}{W} \frac{\pi\alpha\theta_{\omega}}{(\sqrt{\kappa} +1)^2}\frac{I_{\omega, q}^{in}\cos qx}{(q^2 + Q_{GL}^2)},
\end{equation}
 Considering that 
the fraction of the electrons injected from the emitter and crossed the PGL is equal $(1 - p)$ 
and that
 the  spatially uniform and nonuniform components  of the electron current density emitted from the PGL are, respectively,  given by
 
 \begin{equation}\label{eq10}
j_{GL,0} = \frac{4\pi\,e\,\alpha\theta_{\omega}}{(\sqrt{\kappa} +1)^2}I_{\omega, 0}^{in},
\end{equation}
  
\begin{equation}\label{eq11}
j_{GL,q} = \frac{4\pi\,e\,\alpha\theta_{\omega}}{(\sqrt{\kappa} +1)^2}\,I_{\omega, q}^{in}\cos qx,
\end{equation}
 for the components of the electron photocurrent density injected to the p$^+$-layer, one can obtain

 \begin{equation}\label{eq12}
j_{C,0} = \frac{4\pi\,e\,\alpha\theta_{\omega}}{(\sqrt{\kappa} +1)^2}\biggl(\frac{1 - p}{p}
 + 1\biggr)\,
I_{\omega, 0}^{in},
\end{equation}

 \begin{equation}\label{eq13}
j_{C,q} = \frac{4\pi\,e\,\alpha\theta_{\omega}}{(\sqrt{\kappa} +1)^2}\biggl[\frac{1 - p}{p}\frac{Q_{GL}^2}{(q^2 + Q_{GL}^2)}
 + 1\biggr]
I_{\omega, q}^{in}\cos qx.
\end{equation}

 The first term in the brackets in Eqs.~(12) and (13) are due to the contribution of the photoelectric gain effect.
 When the GL lateral conductivity increased, the parameter $Q_{GL}^2$ tends to zero, so that the photoelectric gain effect for the nonuniform current vanishes. 
 
If $Q_{GL}$ tends to zero, Eqs.~(12) and (13) become similar to  the pertinent equation in Ref.~\cite{12}.
Some distinctions are associated with different photosensitivity  of the emitter contacts.

The effect of photoconductive gain becomes substantial when  $Q_{GL}^2/q^2 \gg 1$. 
Depending on the emitter differential conductance, capture probability, GL lateral mobility,
the parameter $Q_{GL}^2$ can vary in a wide range.  
Let us estimate the ratio  $Q_{GL}/q$   for 
$q^{max} = 2\pi/\lambda_{\omega} = 6\pi\times 10^3$~cm$^{-1}$, corresponding to the FIR/MIR with the wavelength $\lambda_{\omega} = 10~\mu$m.    

Using Eqs.~(12) and (13), the photocurrent densities $j_{C,0}$ and $j_{C,q}$ can be expressed via the PGLIP characteristic responsivity
 
 \begin{equation}\label{eq14}
R_{\omega}^{GLIP} = \frac{4\pi\,e\,\alpha\theta_{\omega}}{\hbar\omega(\sqrt{\kappa} +1)^2}.
 \end{equation}
 This yields
  
 \begin{equation}\label{eq15}
j_{C,0} = R_{\omega}^{GLIP}\frac{\hbar\omega\,I_{\omega,0}^{in}}{p},
\end{equation}

 \begin{equation}\label{eq16}
j_{C,q} = R_{\omega}^{GLIP}\biggl[\frac{1 - p}{p}\frac{Q_{GL}^2}{(q^2 + Q_{GL}^2)}
 + 1\biggr]
\hbar\omega\,I_{\omega, q}^{in}\cos qx.
\end{equation}

If  $\Delta_{E} = 0.1$~eV, $\Delta_{GL} = 0.2$~eV,  $j_E^{max} = j_{GL}^{max} = 1.6\times 10^6$~A/cm$^2$,$m =0.3m_0$ ($m_0$ is the mass of bare electron), and $p = 10^{-2}$, one obtains $E_{E}^{tunn} \simeq 2\times 10^6$~V/cm, $E_{GL}^{tunn} \simeq 5.66\times 10^6$~V/cm,  $E_E = E_C = V_0/2W \simeq 0.795\times 10^6$~V/cm,
and  $\sigma_E \simeq 0.41$~A/V$\cdot$cm. At  $\Delta_{E} = 0.2$~eV and  $\Delta_{GL} = 0.4$~eV, one obtains
$E_{E}^{tunn} \simeq 5.66\times 10^6$~V/cm, $E_{GL}^{tunn} \simeq 16\times 10^6$~V/cm,  $E_E = E_C = V_0/2W \simeq 1.84\times 10^6$~V/cm, and $\sigma_E \simeq 0.12$~A/V$\cdot$cm. Using these data, setting $W = 10^{-6}$~cm and $\rho_{GL} > 5$~k$\Omega$, we find $Q_{GL} \gtrsim 5.0\times 10^3$~cm$^{-1}$ and $Q_{GL} \gtrsim 2.7\times 10^3$~cm$^{-1}$, respectively. This implies that to achieve the ratio $Q_{GL}/q \gg 1$ for $q$, corresponding to the incident FIR with the wavelength $\lambda_{\omega} = 10~\mu$m, one needs to use the GLs with the resistivity
much larger than $5$~k$\Omega$.

The radiative recombination of the electrons injected to the LED p-layer with the holes provides the emission
of NIR/VIR photons with the energy $\hbar\Omega > \Delta_G$, where $\Delta_G$ is the energy gap of the p-layer.
The intensity of the output NIR/VIR stimulated by the incident FIR/MIR 
$I_{\omega}^{out} = I_{\Omega, 0}^{out} + I_{\Omega, q}\cos qx$ is determined by the internal quantum efficiency $\tau_n/(\tau_n + \tau_{rad})$  (where $\tau_n$ and $\tau_{rad}$ are the times of nonradiative and radiative recombination, respectively) and by   the fraction  of the generated NIR/VIR photons trapped in  the LED  p-layer due to total internal reflection $\eta$.  

Considering the electron diffusion in the p$^+$-LED layer and the effect of recycling of the NIR/VIR photons~\cite{49,50,51,52,53} in this layers, the density of the electrons  produced by the photocurrent $\Sigma_{LED}$, which  comprises the uniform and spatially nonuniform components, can be found as in Refs.~\cite{8,52,53}:

 \begin{equation}\label{eq17}
\Sigma_{LED,0} =  \frac{j_{C,0}}{e\displaystyle\biggl(\frac{1}{\tau_n} + \frac{1 - \eta}{\tau_{rad}}\biggr)},
\end{equation}

 \begin{equation}\label{eq18}
\Sigma_{LED,q} = \frac{j_{C,q}}{e\displaystyle\biggl[\frac{1}{\tau_n} + \frac{1 - \eta}{\tau_{rad}} + 
 \frac{q^2}{\tau_{rad}}\biggl(\frac{\eta}{q^2 + \ae^2} + L_D^2\biggr)\biggr]}.
\end{equation}
Here  $\ae$ and $L_D = \sqrt{D\tau_{rad}}$ are the interband absorption coefficient of the NIR/VIR photons and the electron diffusion length in the LED  p$^+$-layer.
Using Eqs.~(17) - (18), we obtain

\begin{equation}\label{eq19}
I_{\Omega,0}^{out} =  \frac{j_{C,0}}{e\displaystyle\biggl(\frac{1}{\tau_n} + \frac{1 - \eta}{\tau_{rad}}\biggr)}\frac{\Theta^{out}(1 - \eta)}{\tau_{rad}},
\end{equation}

 \begin{eqnarray}\label{eq20}
I_{\Omega, q}^{out} = \frac{j_{C,q}}{e\displaystyle\biggl[\frac{1}{\tau_n} + \frac{1 - \eta}{\tau_r} +
 \frac{q^2}{\tau_r}\biggl(\frac{\eta}{ q^2 + \ae^2} + L_D^2\biggr)\biggr]} \nonumber\\
 \times\frac{\Theta^{out}(1 - \eta)}{\tau_{rad}}.
\end{eqnarray}
Here $\Theta^{out} \leq 1$ characterizes the ratio of the NIR/VIR photons leaving the LED out of it and those entering to the GLIP ($\Theta^{out}$  depends on the ratios of the refractive indices of the LED p-layer and the 
surrounding layers.

\section{Derivation of up-conversion characteristics}

Substituting $j_{C,0}$ and $j_{C,q}$ from Eqs.~(15) and (16) to Eqs.~(19) and (20), for the pixelless PGLIP-LED average up-conversion and image up-conversion efficiencies defined as
$$
C^{GLIP-LED}_{\omega \rightarrow \Omega, 0} =
\frac{\Omega}{\omega}\frac{I_{\Omega, 0}^{out}}{I_{\omega,0}^{in}}, \qquad C^{GLIP-LED}_{\omega \rightarrow \Omega, q} =
\frac{\Omega}{\omega}\frac{I_{\Omega, q}^{out}}{I_{\omega,q}^{in}},
$$
respectively, we arrive at the following formulas:

\begin{equation}\label{eq21}
C^{GLIP-LED}_{\omega \rightarrow \Omega, 0}=  \frac{\hbar\Omega\Gamma}{e}\frac{R_{\omega}^{GLIP}}{p},
\end{equation}

 \begin{equation}\label{eq22}
C_{\omega\rightarrow\Omega,q}^{GLIP-LED}  = \frac{\hbar\Omega\Gamma}{e}\frac{R_{\omega}^{GLIP}F_{ q}^{LED}}
{P_{q}^{GLIP}}.
\end{equation}
Here

\begin{equation}\label{eq23}
\Gamma = \frac{\Theta^{out}(1-\eta)}{(\tau_{rad}/\tau_n + 1 - \eta)},
\end{equation}

\begin{equation}\label{eq24}
\frac{1}{ P_{q}^{GLIP}} = 1 + \displaystyle\frac{1 - p}{p}\frac{Q_{GL}^2}{(q^2 + Q_{GL}^2)},
\end{equation}
\begin{equation}\label{eq25}
F_{ q}^{LED} = \frac{1}{1 + \displaystyle\frac{q^2}{(\tau_{rad}/\tau_n + 1 - \eta)}\biggl(\frac{\eta}{q^2 + \ae^2} + L_D^2\biggr)}.
\end{equation}
The function $1/ P_{q}^{GLIP}$ describes the role of the GL lateral potential spreading with decreasing of the nonuniformity scale.

The image contrast transfer function, i.e.,
 the ratio of the conversion efficiencies of the image nonuniform component to its  averaged value
 $$
K _{\omega\rightarrow\Omega,q}^{GLIP-LED} = \frac{C_{\omega\rightarrow\Omega,q}^{PGLIP-LED}}{C^{PGLIP-LED}_{\omega \rightarrow \Omega, 0}},
$$ 
which characterizes the output NIR/VIR image contrast, as follows from Eqs.~(21) and (22) is described 
by

\begin{eqnarray}\label{eq26}
K_{\omega\rightarrow\Omega,q}^{GLIP-LED}  = \frac{p}{P_{q}^{GLIP}}F_{q}^{LED}\nonumber\\
= \frac{p + (1 - p)\displaystyle\frac{Q_{GL}^2}{(q^2 + Q_{GL}^2)}}{1 + \displaystyle\frac{q^2}{(\tau_{rad}/\tau_n + 1 - \eta)}\biggl(\frac{\eta}{q^2 + \ae^2} + L_D^2\biggr)}.
\end{eqnarray}

\section{Results}

\begin{figure}[t]
\centering
\includegraphics[width=6.0cm]{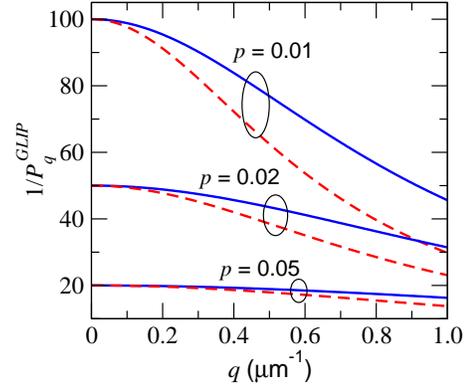}
\caption{PGLIP potential spreading factor  $1/P_{q}^{GLIP}$ versus image nonuniformity wave number $q$
 for different capture probabilities $p$ and   PGL resistivity $\rho_{GL} = 20$~k$\Omega$ (solid lines)  and 10~~k$\Omega$ (dashed lines).}
\end{figure}

\begin{figure}[t]
\centering
\includegraphics[width=8.0cm]{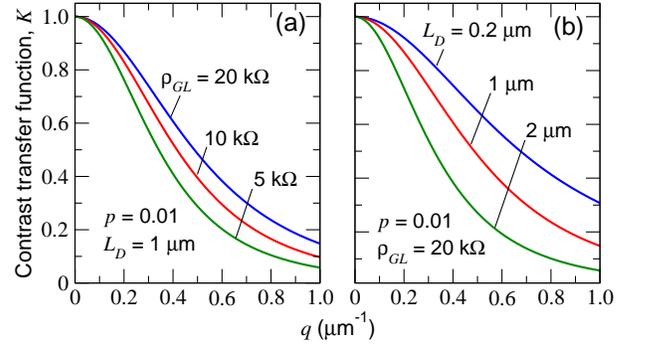}
\caption{Image contrast transfer function
$K _{\omega\rightarrow\Omega,q}^{GLIP-LED}$ versus
image nonuniformity wave number $q$ 
for (a)  different PGL resitivities $\rho_{GL}$ and electron diffusion length in the LED p-layer  $L_D = 1.0~\mu$m  and (b) different  $L_D$ 
  and  $\rho_{GL} = 20$~k$\Omega$. 
}
\end{figure}

 \begin{figure}[t]
\centering
\includegraphics[width=6.0cm]{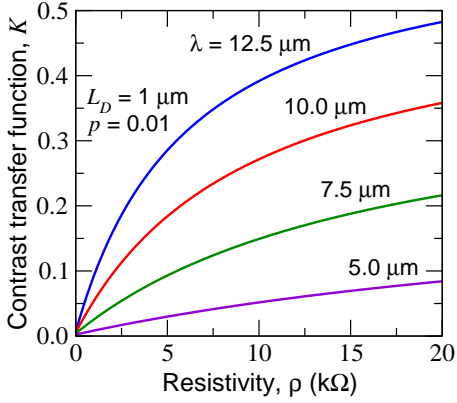}
\caption{ Image contrast transfer function
$K _{\omega\rightarrow\Omega,q}^{GLIP-LED}$ as a function of the PGL resistivity
for different NIR/MIR wavelength $\lambda=2\pi/q$ (5, 7.5, 10, and 12.5 $\mu$m).
}
\end{figure}

 \begin{figure}[t]
\centering
\includegraphics[width=6.0cm]{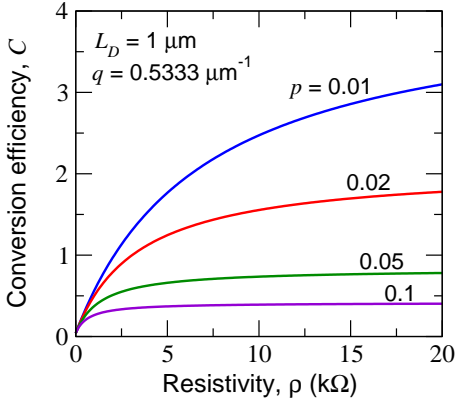}
\caption{ Image conversion efficiency
$C _{\omega\rightarrow\Omega,q}^{GLIP-LED}$ as a function of the PGL resistivity
for different capture probabilities $p$,   $q = q_{max} = \omega/c$,
$\hbar\omega = 0.1$~eV , $\hbar\Omega = 1.0$~eV.
}
\end{figure}

The plots based on the above calculations of the up-conversion characteristics are shown in Figs.~3 - 6. For the definiteness, the following device parameters are assumed:   and $\sigma_E = 0.41$~A/V$\cdot$cm, $\kappa = 5$,
$W = 10^{-6}$~cm, $\tau_{rad}/\tau_n = 0.1$, and $\eta = 0.5$.
Other parameters are indicated below. 

Figure~3 shows the dependence of the  the PGL lateral potential spreading factor $1/P_{q}^{GLIP}$ on the image wavenumber $q$ calculated
using Eq.~(24) for different values of the capture probability $p$ and the PGL resistivity $\rho_{GL}$. One can see that a decrease in the capture probability $p$
(leading to an increase of the photoconductive gain and, therefore, in the enhancement of the role of the nonuniform injection from the emitter), can markedly 
suppress the lateral potential spreading. An increase in the PGL resistivity also
promotes the latter (compare the solid and dashed lines in Fig.~3).

Figure~4 shows the image contract transfer function   
$K _{\omega\rightarrow\Omega,q}^{GLIP-LED}$ versus  
  the image wavenumber $q$ calculated for  different PGL resitivities $\rho_{GL}$ and electron diffusion length in the LED p-layer  $L_D$.
  The pertinent calculations are based on Eqs.~(22) -(26). The following parameters are assumed:
  $p = 0.01$,   $\tau_{rad}/\tau_n = 0.1$, and $\ae = 1~\mu$m$^{-1}$.
 In particular, Fig.~4(a) indicates an improvement of the contrast of the NIR/VIR image when the PGL resistivity rises. This is due to the pertinent suppression of the PGL potential spreading shown in Fig.~3. As follows from Fig.~4(b), a weaker lateral diffusion of the electrons injected into the LED p$^+$-layer   also promotes better
contrast.  

Plots of  $K _{\omega\rightarrow\Omega,q}^{GLIP-LED}$
and  $C_{\omega\rightarrow\Omega,q}^{GLIP-LED} $ shown in 
Figs.~5 and 6 indicate that using  the PGLs with higher resistivity  improves the output image contrast and increases the energy conversion efficiency. 
Diminishing the photon recycling effect in the LED (characterized by parameters $\eta$ and $\ae$ in Eqs.~(25) and (26)), which attenuates the effective electron diffusion,also leads to a higher quality of the  output image.

 Figure~6 
corresponds to  
$\hbar\omega = 0.1$~eV, $\hbar\Omega = 1.0$~eV, $q =  q_{max}= \omega/c = 0.5333~\mu$m$^{-1}$,  $\Gamma = 0.5$, $\theta_{\omega} = 0.5$, and other parameters
 as in the above figures.
 As seen from Fig.~6, the energy conversion efficiency can markedly exceed unity.
 The optimization of the parameters, first of all the LED parameters (increase in the internal and external LED efficiencies), might provide even higher values of the PGLIP-LED energy conversion efficience than those in Fig.~6.

As follows from Figs.~4(a), 5, and 6, the PGLIP-LED up-converter characteristics improve with increasing PGL lateral resistivity $\rho_{GL}$. 
This is attributed to the increasing role of the photoconductive gain when the resistivity rises (see below).   Using PGLs with small
grain sizes $a$~\cite{41,42} (see also Refs.~\cite{43,44,45,46,47}), one can realize the resistivities much higher than those considered in the above figures (say, $\rho_GL > $500~k$\Omega$ at $a \sim 1$~nm).

As mentioned in Sec.~II, a deviation of $|V - V_0|$ from zero leads to  violation of condition (1) and, hence, to a decrease in the GL responsivity $\rho_{GL}$ due to the deviation of the Fermi level in the  GL from
the Dirac point. A decrease in the resistivity with increasing carrier density in the PGL is complicated by
the mobility density dependence and the specifics of the inter-grain transport.
Deviation of the voltage from the that corresponding to the Dirac (neutrality) point in rather wide range might lead to a decrease of the PGL  resistivity by several times\cite{41}. 
This, can result in a marked decrease in the contrast transfer function and the conversion efficiency (see Figs.~5 and 6, respectively).

\section{Materials for PGLIP-LED devices}

Different materials can be used for the PGLIP-LED layered structures, providing their proper relations between
the electron affinities~(see, in particular,~Refs.\cite{54,55}). 

For example, the PGLIP section can include:\\ 
(a) n-Si emitter, SiO$_2$ or hBN -emitter barrier, WS$_2$ collector barrier (as in GL-based vertical-field effect
transistors~\cite{22});\\
  (b) the  n-Si emitter, Si0$_2$ emitter barrier layer (as in GL-based hot electron transistors~\cite{57,58,59})
and  Si collector barrier layer;\\
(c) the  Ti-base emitter, Al$_2$O$_3$ emitter barrier layer, and Si  collector barrier layer  
(the material of the LED active layer  should have the electron affinity  and  energy gap larger than that in Si);\\
(d) As an option,
the emitter layer can also be an n-type GL (as considered in Refs.~\cite{12,13,14}).

The LED active (emitting NIR/VIR) layer can, in particular, be made of such a direct bandgap material as WS$_2$, WSe$_2$, MoSe$_2$, MoS$_2$ ~\cite{28,29,30}. In particular, in the case of the WS$_2$ and MoSe$_2$ LED active layers, the energy of the output image photons is in the range
 $\hbar\Omega \sim 1.5 -1.7$~eV (depending on the temperature).

 \section{Discussion}

\subsection{Effect of photoconductive gain}

The electron photoemission from the PGL leads not only to the photocurrent generation but also (due to the PGL charging and the consequent variation of its potential) to the injection
of extra electrons from the emitter. The latter results in a higher net photocurrent in comparison with the photocurrent provided  solely by the photoemission from the PGL that constitutes what is usually called as the 
effect of the photoconductive gain. 
The photoconductive gain is described by the factor $1/P^{GLIP}_q$ in Eqs.~(22) and (26). As follows from this factor definition given by   Eq.~(24), at $\rho_{GL}$ tending to zero
the factor  $1/P^{GLIP}_q$ tends to unity. This is because at  small values of $\rho_{GL}$, the PGL is virtually equipotential.
This implies that in the  limit $\rho_{GL} = 0$,  the photoconductive gain of the nonuniform photocurrent component vanishes,  so that  the nonuniformity of the  photocurrent injected to the LED p$^+$ layer and, hence,  the nonuniformity of the output radiation intensity   are associated
only with the electrons photoexcited from the PGL. 
Simultaneously,
the uniform (averaged) component still can exhibit a substantial gain, i.e., such a component comprises not only the photocurrent created by the electrons photoexcited  from  the PGL but also by the photocurrent associated with the extra electrons injected from the emitter. This, in particular, seen from Eqs.~(21), (22), and (24), where $C^{GLIP-LED}_{\omega \rightarrow \Omega,0} \propto 1/p \gg 1$, while 
$C^{GLIP-LED}_{\omega \rightarrow \Omega,q}$ does not contain a large factor $1/p$
(in the limit $\rho_{GL} = 0$).  As a consequence, at  small values of  $\rho_{GL}$, the contrast transfer function becomes very small
($K^{GLIP-LED}_{\omega \rightarrow \Omega,0} \simeq p \ll 1$) as it seen in Figs.~5 and 6.
The same happens in the QWIP-LED image up-converters with a single QW, because this QW inevitably must be highly conducting (i.e., have a small QW resistivity $\rho_{QW}$) to provide the carrier density sufficient for a reasonable photosensitivity.

In the multiple-PGL devices,
the spatially uniform and nonuniform components of the photocurrent output from the PGLIP with the resistive PGLs are virtually independent of the number of the PGLs $N$. Hence,
in such  multiple-PGL devices
with  all resistive PGLs,  the photoelectric gain and almost all PGLIP-LED characteristics are  close to those of  the PGLIPs with a single PGL. However, the GLIP-LED image up-converters can exhibit lower noise 
(by a factor of $1/\sqrt{N}$ (see, for example, Refs.~\cite{34}). 

\subsection{PGLIP-LED versus QWIP-LED}

Comparing the image up-conversion efficiency of the PGLIP-LEDs with a single PGL, given by Eqs.~(21) - (24),
with that of the QWIP-LED imagers~\cite{5,8} (assuming the same properties of the LED sections), we find 
\begin{eqnarray}\label{eq27}
\frac{C^{GLIP-LED}_{\omega \rightarrow \Omega,q}}{C^{QWIP-LED}_{\omega \rightarrow \Omega,q}} = 
\biggl(\frac{\alpha}{\sigma_{GL}\Sigma_{GL}}\biggr)\frac{p_{QW}}{[1 -(1-p_{QW})^N]}\nonumber\\
\times  \biggl[1 + \displaystyle\frac{1 - p}{p}\frac{Q_{GL}^2}{(q^2 + Q_{GL}^2)}\biggr]
\simeq \biggl(\frac{\alpha}{\sigma_{GL}\Sigma_{GL}}\biggr)\frac{1}{pN}.
\end{eqnarray}
Here $\sigma_{GL}$, $\Sigma_{QW}$,  $p_{QW}$, and $N$ are the cross-section of the photon absorption and 
the electron density  in the QW, the electron capture probability onto the QW, and the number of the QWs in the QWIP. It is assumed for  simplicity  that the LED sections of both image up-converters have the same characteristics, the number of the QWs is not too large (say, several thens or less), the GL resistivity and  the scale of the image nonuniformities are sufficiently large $q < Q_{GL},\, 1/l_D$, where $l_D = \sqrt{2WD_B/v_B}$,  $D_B$ and $v_B$ are
the electron diffusion length,  electron diffusion coefficient,  and drift velocity in the barrier layers.
Both factors in the right-hand side of Eq.~(27) are large or very large.

Analogously, for the ratio of 
the image contrast transfer functions of the PGLIP-LED and QWIP-LED  one obtains

\begin{equation}\label{eq28}
\frac{K^{GLIP-LED}_{\omega \rightarrow \Omega,q}}{K^{QWIP-LED}_{\omega \rightarrow \Omega,q}} \simeq \frac{1}{p_{QW}N}.
\end{equation}

Due to small values of the capture probability $p_{QW}$ even at a relatively large  but practical number of the QWs, the ratio $K^{GLIP-LED}_{\omega \rightarrow \Omega,q}/K^{QWIP-LED}_{\omega \rightarrow \Omega,q}$ exceeds unity.

\subsection{Role of lateral diffusion of the injected electrons in the barrier layers}
In Eqs.~(8) and (27) we disregarded the lateral diffusion of the electrons propagating above the barriers (in contrast to Ref.~\cite{5,8}). This is justified because the lateral displacement of these electrons during their rather short flight across the barrier layers is very small. Indeed, such a displacement $\Delta x \simeq L_D$. Setting $W = 10^{-6}$~cm,
$D_B = (10 -100)$~cm$^2$/s, and $v_B = 10^7$~cm/s, we obtain $\Delta x \simeq (1.4 - 4.5)\times 10^{-6}$~cm, i.e.,
the value negligibly small in comparison with the incident radiation wavelength.

\subsection{Role of the GL and barriers  doping (electrical and chemical)}

When the bias voltage $V$ deviates from the characteristic voltage $V_0$, the Fermi energy in the PGL
shifts with respect to the Dirac point. This leads to the following consequences. First, an increase in the electron or hole density $\Sigma_{GL}$ results in the increase in the GL lateral conductivity and, hence, in
the smoothening of the lateral potential distribution and the suppression of the photoelectric gain.
The same occurs when the GL is chemically doped.

Second, the Fermi energy shift  affects the PGL absorption spectrum due the Pauli principle (toward higher 
energies of the FIR/MIR photons~\cite{15}). This might be used for a voltage control of the spectral characteristics (say, for the "filtering") of the incident FIR/MIR.

Third, the deviation of $V$ from $V_0$ as well as chemical doping (giving rise to the formation of the hole gas in the GLIP) can be used for a lowering of the GLIP dark current and, therefore for decrease in the background uniform component of the output NIR/VIR. 

Forth, the selective dipole  doping of the barrier layers can markedly modify the PGLIP characteristics~\cite{15} affecting
the operation of both the  lamp GLIP-LED and PGL-LED  up-converters~\cite{12} and  the pixelless PGLIP-LED imagers.

\subsection{Optical feedback}

If a substantial portion of the NIR/VIR photons generated in the LED active p$^+$-layer (and not reflected by the PGLIP collector barrier)  enters the
PGLIP (the pertinent wavy arrows are not shown in Figs.~1 and 2), the interband absorption of these photons in the PGLIP (in the GL) leads to an extra photocurrent, which further reinforces the emission of the NIR/VIR photons. Such a positive optical feedback can reinforce not only the average up-conversion efficiency~\cite{12}  but the image up-conversion as well.

\section{Conclusions}

We reported on the proposal of  the pixelless FIR/MIR to NIR/VIR   up-converter
  based on the vdW heterostructures with the highly resistive (polycrystalline) PGLs
 -  PGLIP-LED upconverter. 
Using the  developed device model which accounts for generation of nonuniform photocurrent in the GLIP section by FIR/MIR, the photocurrent injection to the LED section, and emission of NIR/VIR from the latter section, we calculated the PGLIP-LED characteristics (the image contrast transfer function and the conversion efficiency). 
The photocurrent lateral spreading was considered  taking into account the PGL lateral conductivity and the effective  diffusion 
of the electrons injected into LED (combining  their standard diffusion and the lateral spreading due to the photon recycling).
 We showed that the pixelless PGLIP-LED up-converters can be effective imaging devices exhibiting the power image up-conversion efficiency substantially exceeding unity. Recent publications~\cite{54,55,56,57} and others support the feasibility of
 realization of the PGLIP-LED devices with elevated performance. The proposed and evaluated pixelless PGLIP-LED up-converters can markedly surpass the
 pixelless QWIP-LED imagers.

\section*{Acknowledgments}

The authors are grateful to D. Svintsov, V. Leiman, V. Mitin, and A. Satou for useful information and numerous discussions. The work at RIEC and UoA was supported by the Japan Society for Promotion of Science, KAKENHI Grant No. 16H06361. The work at BMSTU was supported  
 by the Russian Scientific Foundation, Grant No. 14-29-00277. VR  and MS acknowledge the support 
 by the Russian Foundation for Basic Research, Grant  
No.16-37-60110/16 and  by the US ARL Cooperative Research Agreement, respectively.

 \end{document}